\documentclass[aps,pra,twocolumn,superscriptaddress,floatfix,
nofootinbib,showpacs,longbibliography]{revtex4-1}

\usepackage[utf8]{inputenc}  
\usepackage[T1]{fontenc}     
\usepackage[british]{babel}  
\usepackage[sc,osf]{mathpazo}
\usepackage[scaled=0.86]{berasans}  
\usepackage[colorlinks=true, citecolor=blue, urlcolor=blue]{hyperref}  
\usepackage{graphicx} 
\usepackage[babel]{microtype}  
\usepackage{amsmath,amssymb,amsthm,bm,amsfonts,mathrsfs,bbm} 

\usepackage{xspace}  
\usepackage{xcolor}
\usepackage{multirow}
\usepackage{array}
\usepackage{bigstrut}
\usepackage{braket}
\usepackage{color}
\usepackage{multirow}
\usepackage{mathtools}
\usepackage{float}
\usepackage[caption = false]{subfig}
\usepackage{xcolor,colortbl}
\usepackage{color}

\newcommand{\be}{\begin{equation}}
\newcommand{\ee}{\end{equation}}
\newcommand{\ba}{\begin{eqnarray}}
\newcommand{\ea}{\end{eqnarray}}

\newcommand{\toph}[3]{\ket{#1}_A\ket{#2}_B\ket{#3}_C}

\newtheorem{definition}{Definition}
\newtheorem{proposition}{Proposition}

\newtheorem{remark}{Remark}

	\definecolor{rvwvcq}{rgb}{0,0,1}

\def\>{\rangle}
\def\<{\langle}

\newcommand{\bi}[2]{\ket{#1}\ket{#2}}
\newcommand{\tri}[3]{\ket{#1}\ket{#2}\ket{#3}}

\begin{document}

\title{Genuinely Nonlocal Product Bases: Classification and Entanglement Assisted Discrimination}

\author{Sumit Rout}
\affiliation{Integrated Science Education and Research Centre, Visva Bharati University, Santiniketan 731235, West Bengal, India.}

\author{Ananda G. Maity}
\affiliation{S.N. Bose National Center for Basic Sciences, Block JD, Sector III, Salt Lake, Kolkata 700106, India.}

\author{Amit Mukherjee}
\affiliation{Optics and Quantum Information Group, The Institute of Mathematical Sciences,	HBNI, C. I. T. Campus, Taramani, Chennai 600113, India.}

\author{Saronath Halder}
\affiliation{Department of Mathematics, Indian Institute of Science Education and Research Berhampur, Transit Campus, Government ITI, Berhampur 760010, India.} 

\author{Manik Banik}
\affiliation{S.N. Bose National Center for Basic Sciences, Block JD, Sector III, Salt Lake, Kolkata 700098, India.}

\begin{abstract}
An orthogonal product basis of a composite Hilbert space is {\it genuinely nonlocal} if the basis states are locally indistinguishable across every bipartition. From an operational point of view such a basis corresponds to a separable measurement that cannot be implemented by local operations and classical communication (LOCC) unless all the parties come together in a single location. In this work we classify genuinely nonlocal product bases into different categories. Our classification is based on state elimination property of the set via orthogonality-preserving measurements when all the parties are spatially separated or different subsets of the parties come together. We then study local state discrimination protocols for several such bases with additional entangled resources shared among the parties. Apart from consuming less entanglement than teleportation based schemes our protocols indicate operational significance of the proposed classification and exhibit nontrivial use of genuine entanglement in local state discrimination problem.
\end{abstract}



\maketitle
\section{Introduction}
Superposition principle lies at the core of quantum mechanics which leads to several no-go results in quantum information theory, such as, no-cloning \cite{Wootters82} and no-deleting theorem \cite{Pati04}. It also gives rise to the concept of nonorthogonal states for which perfect discrimination is never possible. Origin of the quantum state discrimination problem dates back to early nineteen seventies with an initial attempt to formalize information processing with optical quantum devices \cite{Helstrom69,Holevo73,Yuen75}. Although sets of mutually orthogonal states can always be perfectly distinguished by some global measurements, the situation may change dramatically for a set of such multipartite quantum states if the spatially separated parties are allowed to perform only local operations assisted with classical communication (LOCC). In a seminal work Bennett {\it et al.} provided examples of mutually orthogonal product states that are indistinguishable under LOCC given one copy of the state \cite{Bennett99}. They coined the term `quantum nonlocality without entanglement' \footnote{Note that this nonlocal feature is different than the concept of `quantum nonlocality' as established in another seminal work by John S. Bell \cite{Bell64}. A multipartite input-output correlation is called nonlocal if it is not compatible with the classical description of {\it local-realism} (see \cite{Brunner14} for a review on Bell nonlocality). In quantum world such correlations can only be resulted from multipartite entangled states.} for this phenomenon as the states allow local preparation (with some preshared strategy) but prohibit perfect local discrimination. Importantly, local indistinguishability turns out to be a crucial primitive for a number of distributed quantum protocols, namely, quantum data hiding \cite{Terhal01,Eggeling02} and quantum secret sharing \cite{Markham08,Rahaman15,Wang17-0}.              

The result of Bennett {\it et al.} \cite{Bennett99} motivates overwhelming research interest on generic local state discrimination problems — the task of optimal discrimination of
multiparty states, not necessarily product, by means of LOCC \cite{Bennett99-1,Walgate00,Virmani01,Ghosh01,Groisman01,Walgate02,DiVincenzo03,Horodecki03,Fan04,Ghosh04,Rinaldis04,Nathanson05,Watrous05,Niset06,Ye07,Fan07,Duan07,Bandyopadhyay09,Feng09,Duan10,Yu12,Yang13,Childs13,Zhang14,Yu15,Zhang15,Wang15,Chen15,Yang15,Zhang16,Xu16,Zhang16-1,Xu16-1,Wang17,Zhang17,Xu17,Wang17-1,Zhang17-1,Zhang17-2,Zhang17-3,Corke17,Halder18}. Very recently, Halder {\it et al} have introduced a nontrivial generalization of the quantum nonlocality without entanglement phenomena \cite{Halder19}. They have provided examples of $3$-qutrit and $3$-ququad product bases that are not distinguishable even if (any) two of the three parties come together, {\it i.e.}, each of these product basis can be prepared locally but to distinguish them either all the three parties need to come together or entangled resources must be shared across all bipartitions. Quite naturally, one can define such a feature as {\it genuine quantum nonlocality without entanglement}. Existence of the product bases of Ref. \cite{Halder19} has important operational consequences. While in one hand they constitute a nontrivial primitive for multipartite information theoretic protocols, on the other hand they correspond to multipartite separable measurements that require entanglement resources across all bipartitions for implementing those measurements.

In this work, we classify genuinely nonlocal product bases (GNPBs) of multipartite quantum systems. The classification is based on whether such a basis is reducible, {\it i.e}, allows elimination of state(s) from the set under orthogonality-preserving measurements (OPMs) when all the parties are spatially separated or some subset of the parties are allowed to come together. We then show that this way of classification has interesting operational consequences. We provide example of multipartite GNPBs that allow local elimination of some states under OPM even when all the parties are separated and subsequently they require entanglement across one bipartite cut only for perfect discrimination of the states. In other words, local elimination makes it adequate to consume less entanglement for perfect discrimination of the states in those GNPBs. We then provide examples of tripartite GNPB which is locally irreducible when all three parties are in separate location but reducible if two of the parties are together. However, it requires entanglement across every bipartite cut for perfect discrimination. Such a GNPB is weaker than the GNPB of \cite{Halder19} as the later does not allow any elimination of states under nontrivial OPM even when any two parties come together. We then provide different entanglement assisted protocols for perfect discrimination of several GNPBs. To the best of our knowledge, the entanglement assisted discrimination of GNPBs that require entanglement across every bipartition for perfect local discrimination is reported in the present manuscript for the very first time. The studied protocols are resource efficient as they consume less entanglement in comparison to the teleportation based protocols. Interestingly, one of our protocols exhibits nontrivial and advantageous use of genuine entanglement in local state discrimination problem.       

We arrange present manuscript in the following way: in Section~\ref{sec2} we briefly discuss the notations and some required preliminary concepts, Sec.~\ref{sec3} \& \ref{sec4} consist of the main contributions of the present work, in Sec.~\ref{sec5} we summarize the results with a discussion of some open problems.

\section{Preliminaries and notations}\label{sec2}
Throughout the paper we will use standard notations and terminologies that are commonly used in quantum information theory. All the systems we consider are finite dimensional and thus, the associated Hilbert spaces are isomorphic to some complex Euclidean spaces $\mathbb{C}^d$, with $d\in\mathbb{N}$ denoting the dimension of the system. Composite quantum systems are associated with tensor product of the corresponding subsystems' Hilbert spaces -- an $n$-partite quantum system is associated with the Hilbert space $\bigotimes_{i=1}^n\mathbb{C}^{d_i}$, where $\mathbb{C}^{d_i}$ corresponds to the $i^{th}$ subsystem. For our purpose we start with recalling the following definition.

\begin{definition}\label{def1}
Nonlocal product bases (NPBs):- Consider an  $n$-partite quantum system with Hilbert space $\bigotimes_{i=1}^n\mathbb{C}^{d_i}$. An orthogonal product basis (OPB), $\mathbb{B}_{nl}\equiv\left\lbrace \ket{\psi}_j=\bigotimes_{i=1}^n\ket{\alpha}^i_j~|~j=1,\cdots,\Pi_{i=1}^nd_i\right\rbrace$ is called nonlocal if the states in $\mathbb{B}_{nl}$ can not be perfectly distinguished by LOCC when all the parties are spatially separated.
\end{definition}  
Bipartite as well as multipartite examples of such bases were first constructed in \cite{Bennett99} for $(\mathbb{C}^3)^{\otimes 2}$ and $(\mathbb{C}^2)^{\otimes 3}$ Hilbert spaces. Though for the second example the states cannot be discriminated under LOCC when all of the three parties are spatially separated, but can be done perfectly when two of the parties come together. This observation leads to a stronger notion of nonlocality without entanglement.
\begin{definition}\label{def2}
Genuinely nonlocal product bases (GNPBs):- A multipartite OPB, $\mathbb{B}_{gnl}\equiv\left\lbrace \ket{\psi}_j=\bigotimes_{i=1}^n\ket{\alpha}^i_j~|~j=1,\cdots,\Pi_{i=1}^nd_i\right\rbrace \subset\bigotimes_{i=1}^n\mathbb{C}^{d_i}$ is called genuinely nonlocal if the states in $\mathbb{B}_{gnl}$ can not be perfectly distinguished by LOCC even if any $(n-1)$ parties are allowed to come together. 
\end{definition}
Ref. \cite{Halder19} provides examples of such product bases for $3$-qutrit and $3$-ququad quantum systems. At this point, let us discuss a bit about local protocols. These are generally multi-round protocols. For multipartite case, depending on the scenarios whether all parties are separated or some group of parties are allowed to come together, in a particular round each party individually and/or some as a group perform(s) local quantum operations and communicate(s) the classical outcomes to other parties and/or other groups. Depending on these communications other parties (and/or groups) further choose their actions and the protocol goes on. Being multi round, it is in general difficult to mathematically characterize the set of LOCC operations. Some interesting topological behaviors of the LOCC set have been identified in  \cite{Chitambar14}. While discriminating a set of mutually orthogonal multipartite product states by such a LOCC protocol, in a given round either the given state must be identified or some of the states must be eliminated. If the later happens then for perfect discrimination the remaining post measurement states should be mutually orthogonal so that the protocol can be further carried on. This leads to the following definition.    
\begin{definition}\label{def3}
Nontrivial- orthogonality preserving measurement (N-OPM):- A measurement performed to distinguish a set of mutually orthogonal quantum states is called orthogonality preserving measurement (OPM) if after the measurement the states remain mutually orthogonal. Furthermore, such a measurement is called nontrivial if all the measurement effects constituting the OPM are not proportional to identity operator, otherwise it is trivial.  
\end{definition}
Definition \ref{def3} subsequently leads to the concept of 'locally irreducible set' --  a set of mutually orthogonal multipartite quantum states from which it is not possible to
eliminate one or more states by orthogonality preserving local measurements. Although local irreducibility sufficiently assures locally indistinguishable but the former is not a necessary requirement for the later one. It turns out that the examples of \cite{Halder19} exhibit a `strong nonlocal' behavior as those product bases are not only GNPBs but they are indeed locally irreducible even if any two parties come together.

While additional entangled states are supplied as resource among the parties along with their operational power LOCC then it may be possible to perfectly distinguish a GNPB. An immediate such protocol follows from quantum teleportation \cite{Bennett93}. If the involved parties share sufficient entanglement so that they can teleport their respective subsystems to one of the parties then she/he can perfectly discriminate the states by performing suitable measurement. Since entanglement is costly resource under the operational paradigm of LOCC, therefore any protocol consuming less entanglement is always desirable. First instance of such protocols for a class of locally indistinguishable product states was proposed by Cohen \cite{Cohen08}. For instance, Bennett's $2$-qutrit NPB can be perfectly distinguished by LOCC with additional $1$-ebit entanglement whereas the teleportation based protocol requires a $2$-qutrit maximally entangled state, {\it i.e.}, $\log3$-ebit. Here, we use the unit ebit, so, logarithm is taken with respect to base $2$. Cohen's result motivates further research in identifying efficient use of entanglement in local state discriminating problem \cite{Bandyopadhyay09-1, Bandyopadhyay10, Duan14, Bandyopadhyay14, Bandyopadhyay16, Zhang16-2, Bandyopadhyay18, Zhang18, Halder18-1, Li19,Zhang18, Halder18-1, Li19}. 

In this work we also study efficient state discrimination protocols for several GNPBs. Note that, in multipartite scenario different types of entangled resource may be supplied as there exist different inequivalent types of entanglement. For instance, in tripartite scenario different pairs of parties can be supplied with $2$-qubit maximally entangled state, {\it i.e.}, Einstein-Podolsky-Rosen (EPR) state, $\ket{\phi^+}:=\frac{1}{\sqrt{2}}(\ket{00}+\ket{11})\in(\mathbb{C}^2)^{\otimes2}$ or they can be supplied with $3$-qubit Greenberger–Horne–Zeilinger (GHZ) state, $\ket{G}:=\frac{1}{\sqrt{2}}(\ket{000}+\ket{111})\in(\mathbb{C}^2)^{\otimes3}$. Here we consider two different configurations of entanglement resources.  

{\bf Config. (1):} $\left\lbrace\left(p,\ket{\phi^+}_{\mathcal{AB}}\right);\left(q,\ket{\phi^+}_{\mathcal{BC}}\right);\left(r,\ket{\phi^+}_{\mathcal{CA}}\right)\right\rbrace$ with $p,q,$ and $r$ taking nonnegative values. 
\begin{figure}[h!]
	\includegraphics[width=0.35\textwidth]{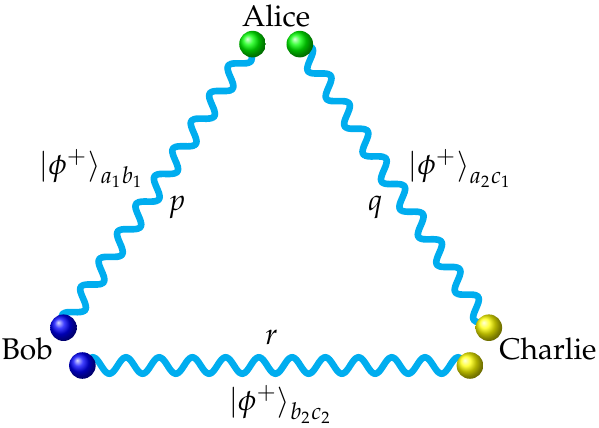}
	\caption{(Color on-line) Spatial configuration of the resource state $\left\lbrace\left(p,\ket{\phi^+}_{\mathcal{AB}}\right);\left(q,\ket{\phi^+}_{\mathcal{BC}}\right);\left(r,\ket{\phi^+}_{\mathcal{CA}}\right)\right\rbrace$. The indices $a_1,a_2\in\mathcal{A}$, $b_1,b_2\in\mathcal{B}$, and $c_1,c_2\in\mathcal{C}$.}
	\label{Fig:3epr}
\end{figure}
It denotes that on average $p$ amount of $2$-qubit maximally entangled state is consumed between Alice \& Bob, and similarly in other two pairs $q$ and $r$ amounts of EPR state are consumed while discriminating a GNPB. Whenever discriminating a GNPB of $(\mathbb{C}^d)^{\otimes3}$, then a successful discrimination protocol with this resource configuration will be efficient than the corresponding teleportation based protocol if $(p+q+r)<2\log d$. 
	 
{\bf Config. (2):} $\left\lbrace\left(p,\ket{G}_{\mathcal{ABC}}\right);\left(q,\ket{\phi^+}_{\star}\right)\right\rbrace$ with $p,q$ taking nonnegative values and $\star$ be one of the pairs from $\{\mathcal{AB},\mathcal{BC},\mathcal{CA}\}$. 
\begin{figure}[h!]
	\includegraphics[width=0.35\textwidth]{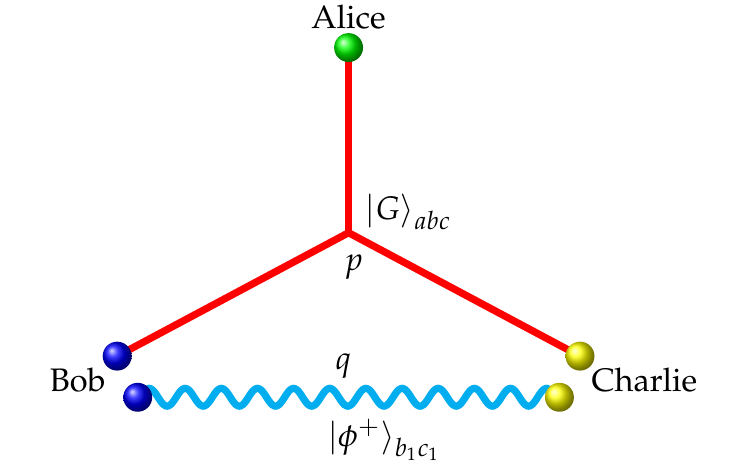}
	\caption{(Color on-line) Spatial configuration of the resource state $\left\lbrace\left(p,\ket{G}_{\mathcal{ABC}}\right);\left(q,\ket{\phi^+}_{\mathcal{BC}}\right)\right\rbrace$. }
	\label{Fig:GHZ}
\end{figure}
It denotes that while discriminating a GNPB, $p$ amount of $3$-qubit GHZ state is consumed in addition with $q$ amount of EPR state shared between one of the three pairs. Note that, to distribute a $3$-qubit GHZ state among Alice, Bob, and Charlie $2$ copies of EPR state shared among two pairs (say) Alice \& Bob and Alice \& Charlie are required -- Alice prepares a GHZ state at her lab and teleports two subsystems to Bob and Charlie respectively. However, the process might be irreversible as there is no known local protocol via which it is possible to get back two copies of two-qubit Bell states from a GHZ state \cite{Bennett00}.  
  
\section{Classification of GNPB(s)}\label{sec3}
The states in an $n$-partite GNPB can not be perfectly distinguished under LOCC even if any $(n-1)$ parties are allowed to come together. However, it may be possible that while some parties come together, they can eliminate some states from the set under local measurement which keeps the post measurement states orthogonal. Based on how many parties are required to come together for such elimination we can classify the GNPBs into different types. In the following we will discuss this classification with explicit examples. Though the classification can be generalized for arbitrary number of parties, we will mainly restrict our study for tripartite Hilbert spaces. 

\subsection{GNPB: Type-I} 
Such a GNPB is locally reducible even when all the parties are separated, {\it i.e.}, some subset of states can be eliminated under nontrivial local OPM. 
  
{\bf Example:} Consider the quantum system with Hilbert space $(\mathbb{C}^4)^{\otimes3}$  shared among Alice, Bob, and Charlie. Computational bases for $\mathbb{C}^4$ will be denoted as $\{\ket{i}\}_{i=0}^3$. We will use the short hand notation $\ket{\alpha}_A\ket{\beta}_B\ket{\gamma}_C$ for the state $\ket{\alpha}_A\otimes\ket{\beta}_B\otimes\ket{\gamma}_C$ and will avoid the party index where possible. To construct the required GNPB, first consider the following set of states: 
\begin{equation}\label{eq1}
\mathcal{S}\equiv\left\lbrace \ket{3}_A\ket{\beta}_{BC},~\ket{\beta}_{AB}\ket{3}_{C}\right\rbrace;
\end{equation}
where $\ket{\beta}$'s are the states belonging in the $2$-qutrit NPB of Ref. \cite{Bennett99}, {\it i.e.}, $\ket{\beta}\in\mathcal{B}\equiv\left\lbrace\bi{0}{\eta_\pm},~\bi{\eta_\pm}{2},~\bi{2}{\xi_\pm},~\bi{\xi_\pm}{0},~\bi{1}{1}\right\rbrace$; where $\ket{\eta_\pm}:=(\ket{0}\pm\ket{1})/\sqrt{2}$ and $\ket{\xi_\pm}:=(\ket{1}\pm\ket{2})/\sqrt{2}$. The definitions of $\ket{\eta_\pm}$ and $\ket{\xi_\pm}$ are maintained same throughout the manuscript.
Also consider the following product states:
\begin{align}\label{c}
\rotatebox[origin=c]{0}{$\mathcal{R}\equiv$}
\left\{\!\begin{aligned}
|3\rangle|0\rangle|3\rangle,~|3\rangle|1\rangle|3\rangle,~
|3\rangle|2\rangle|3\rangle,~|3\rangle|3\rangle|0\rangle,\\
|3\rangle|3\rangle|1\rangle,~|3\rangle|3\rangle|2\rangle,~
|3\rangle|3\rangle|3\rangle,~|2\rangle|0\rangle|0\rangle,\\
|2\rangle|0\rangle|1\rangle,~|2\rangle|0\rangle|2\rangle,~
|2\rangle|1\rangle|0\rangle,~|2\rangle|1\rangle|1\rangle,\\
|2\rangle|1\rangle|2\rangle,~|2\rangle|2\rangle|0\rangle,~
|2\rangle|2\rangle|1\rangle,~|2\rangle|2\rangle|2\rangle,\\
|2\rangle|3\rangle|0\rangle,~|2\rangle|3\rangle|1\rangle,~
|2\rangle|3\rangle|2\rangle,~|2\rangle|3\rangle|3\rangle,\\
|0\rangle|0\rangle|0\rangle,~|0\rangle|0\rangle|1\rangle,~
|0\rangle|0\rangle|2\rangle,~|0\rangle|1\rangle|0\rangle,\\
|0\rangle|1\rangle|1\rangle,~|0\rangle|1\rangle|2\rangle,~
|0\rangle|2\rangle|0\rangle,~|0\rangle|2\rangle|1\rangle,\\
|0\rangle|2\rangle|2\rangle,~|0\rangle|3\rangle|0\rangle,~
|0\rangle|3\rangle|1\rangle,~|0\rangle|3\rangle|2\rangle,\\
|0\rangle|3\rangle|3\rangle,~|1\rangle|0\rangle|0\rangle,~
|1\rangle|0\rangle|1\rangle,~|1\rangle|0\rangle|2\rangle,\\
|1\rangle|1\rangle|0\rangle,~|1\rangle|1\rangle|1\rangle,~
|1\rangle|1\rangle|2\rangle,~|1\rangle|2\rangle|0\rangle,\\
|1\rangle|2\rangle|1\rangle,~|1\rangle|2\rangle|2\rangle,~
|1\rangle|3\rangle|0\rangle,~|1\rangle|3\rangle|1\rangle,\\
|1\rangle|3\rangle|2\rangle,~|1\rangle|3\rangle|3\rangle.~~~~~~~~~~~~~~~~
\end{aligned}\right\},	
\end{align}
These $46$ states in $\mathcal{R}$ along with the states in $\mathcal{S}$ form a OPB in $(\mathbb{C}^4)^{\otimes 3}$. Since the set $\mathcal{B}$ is a NPB in $(\mathbb{C}^3)^{\otimes2}$, therefore the set of tripartite states $\{\ket{3}_A\ket{\beta}_{BC}~|~\ket{\beta}\in\mathcal{B}\}$ are locally indistinguishable across $B|AC$ cut as well as $C|AB$ cut. Similarly the set of states $\{\ket{\beta}_{AB}\ket{3}_C~|~\ket{\beta}\in\mathcal{B}\}$ are locally indistinguishable across $A|BC$ and $B|AC$ cuts. As the considered OPB contains both theses subsets of states, thus we have the following proposition.
\begin{proposition}\label{prop1}
The set of orthogonal states $\mathbb{B}_I(4,3)\equiv\mathcal{S}\cup\mathcal{R}$ is a GNPB in $(\mathbb{C}^4)^{\otimes3}$. However the set is locally reducible even when all the parties are separated. 
\end{proposition} 
Construction of such GNPBs are straightforward for arbitrary number of parties. At this point it is important to note that the local indistinguishability arises due to the {\it twisted} states $\ket{\beta}\in\mathcal{B}$, which are obtained from linear superposition of computational states. In other words, quantum superposition principle plays key role for manifestation of `quantum nonlocality without entanglement' phenomenon. While discriminating the states in $\mathbb{B}_I(4,3)$, any of the parties can perform nontrivial OPM to eliminate certain states even when all of them are separated. For instance, let Alice performs local measurement $M\equiv\{\ket{3}\bra{3},~\mathbb{I}_4-\ket{3}\bra{3}\}$. If the measurement result corresponds to the projector $\ket{3}\bra{3}$, then the given state must be one of the following states:
\begin{align}\label{c12}
\rotatebox[origin=c]{0}{$\ket{3}\Rightarrow$}
\left\{\!\begin{aligned}
\ket{3}\ket{\beta},~|3\rangle|0\rangle|3\rangle,~|3\rangle|1\rangle|3\rangle,~
|3\rangle|2\rangle|3\rangle,\\
|3\rangle|3\rangle|0\rangle,~
|3\rangle|3\rangle|1\rangle,~|3\rangle|3\rangle|2\rangle,~|3\rangle|3\rangle|3\rangle
\end{aligned}\right\}.	
\end{align}
Otherwise it is one of the remaining states. Since the considered measurement is an OPM, so after this step entanglement assisted discrimination protocol can be carried on. The outcome provides nontrivial information in which cut they need to share bipartite entanglement. If the outcome corresponds to the projector $\ket{3}_A\bra{3}$ then entanglement is required to share between Bob \& Charlie, otherwise it should be shared between Alice \& Bob. In other words, local elimination makes it possible to consume entanglement across one bipartite cut only for perfect discrimination of the states. 

\subsection{GNPB: Type-II}
Such a GNPB is locally irreducible when all the parties are separated, {\it i.e.}, no local elimination is possible preserving orthogonality among the post measurement states. 

{\bf Example:} From the GNPB $\mathbb{B}_I(4,3)$ every party can locally eliminate some states by performing OPM that discriminate the subspace spanned by $\ket{3}$ vs subspace spanned by $\{\ket{0},\ket{1},\ket{2}\}$. One will obtain a GNPB of Type-II if this local elimination can be stopped. For this purpose, take the states $\{|0\rangle|3\rangle|2\rangle,~|0\rangle|3\rangle|3\rangle\}\subset\mathbb{B}_I(4,3)$. Consider now a new OPB that contains the locally twisted product 
states $\{|0\rangle|3\rangle|\chi_\pm\rangle\}$ instead of the states $\{|0\rangle|3\rangle|2\rangle,~|0\rangle|3\rangle|3\rangle\}$, where $\ket{\chi_\pm}:=(\ket{2}\pm\ket{3})/\sqrt{2}$. As a consequence Charlie is no more able to discriminate between the subspaces spanned by $\ket{3}$ and $\{\ket{0},\ket{1},\ket{2}\}$ and thus, he cannot eliminate any state via OPM. Similarly, the twisted product states $\{|2\rangle|\chi_\pm\rangle|2\rangle\}$ stop Bob and $\{|\chi_\pm\rangle|3\rangle|1\rangle\}$ stop Alice from eliminating any state via nontrivial OPM. Thus we have the following proposition.
\begin{proposition}\label{prop2}
The set of states,
\begin{align*}\label{t2}
\rotatebox[origin=c]{0}{$\mathbb{B}_{II}(4,3):=$}
\left\{\!\begin{aligned}
\mathbb{B}_{I}(4,3)\setminus\{\tri{0}{3}{2},\tri{0}{3}{3},\\\tri{2}{2}{2},\tri{2}{3}{2},\tri{2}{3}{1},\\\tri{3}{3}{1}\}\\
\cup\{\tri{0}{3}{\chi_\pm},\tri{2}{\chi_\pm}{2},\tri{\chi_\pm}{3}{1}\} 
\end{aligned}\right\},	
\end{align*}
is a GNPB in $(\mathbb{C}^4)^{\otimes3}$. Furthermore the set is locally irreducible when all the parties are specially separated. 
\end{proposition}
Like Type-I basis, in this case also it is also possible to generalize the construction for arbitrary number of parties. Here we find that the set of multipartite nonlocal product states constructed in Ref.~\cite{Zhang17} also posses the similar feature of our Type-II basis. Clearly, the GNPB $\mathbb{B}_{II}(4,3)$ requires entanglement across every bipartition for perfect discrimination. In other words, if two of the parties come together then also the nonlocality persists. Here comes two important observations: (i) local twist plays an important role in the construction of a new class of GNPBs -- it can increase the entanglement cost of distinguishing a product basis by LOCC; (ii) the examples $\mathbb{B}_{I}(4,3)$ and $\mathbb{B}_{II}(4,3)$ also exhibit operational implication of `local elimination via nontrivial OPM' as perfect discrimination of the first set requires entangled resource only in one cut, the later one demands entanglement in more than one cut.

So far we have constructed GNPBs for Hilbert spaces with local subsystem dimension $4$. Naturally the question arises regarding such constructions in lower dimensional cases. Remember that all product bases in $\mathbb{C}^2\otimes\mathbb{C}^d$ are locally distinguishable \cite{DiVincenzo03}. Therefore for a GNPB to exist the minimum dimension is $\mathbb{C}^3\otimes\mathbb{C}^3\otimes\mathbb{C}^3$. However, the technique used above or the technique used in \cite{Zhang17} are not applicable to construct GNPBs in the minimum dimension. In the following we provide an example of GNPB in $(\mathbb{C}^3)^{\otimes3}$.  
\begin{proposition}\label{prop3}
	The set of states,
	\begin{align*}
	\rotatebox[origin=c]{0}{$\mathbb{B}_{II}(3,3)\equiv$}
	\left\{\!\begin{aligned}
	\tri{0}{\eta_\pm}{\xi_\pm},~\tri{\eta_\pm}{2}{\xi_\pm},~\\
	\tri{2}{\xi_\pm}{\eta_\pm},~\tri{\eta_\pm}{\xi_\pm}{0},~\\
	\tri{\xi_\pm}{0}{\eta_\pm},~\tri{\xi_\pm}{\eta_\pm}{2},~\\
	\tri{k}{k}{k}~~|~~ k\in\{0,1,2\}~~~~
	\end{aligned}\right\},	
	\end{align*}
	is a GNPB of Type-II in $(\mathbb{C}^3)^{\otimes3}$.
\end{proposition} 
When the three parties are spatially separated, it is not possible to eliminate any state by OPM from the above set and hence the set is locally indistinguishable. This can be easily proved by the technique described in \cite{Halder19,Halder18-1}. We are yet to prove local indistinguishability of the above set across every bipartitions. For that, first note that set $\mathcal{B}$ is present between any two pair in the above construction. For instance consider the subset of states
\begin{align*}
\left\{\!\begin{aligned}
\tri{\xi_\pm}{0}{\eta_\pm},~\tri{\xi_\pm}{\eta_\pm}{2},~\\
\tri{\eta_\pm}{2}{\xi_\pm},~\tri{\eta_\pm}{\xi_\pm}{0},~\\
\tri{1}{1}{1}~~~~~~~~~~~~~~
\end{aligned}\right\}.	
\end{align*}
Presence of $\mathcal{B}$ between Bob and Charlie is evident here. Furthermore, Alice's states $\ket{\eta_\pm},~\ket{\xi_\pm},$ and $\ket{1}$ are tagged with. These tagged states are not all mutually orthogonal due to the presence of local twist. Moreover, the twisted states saturate the local dimension of Alice, {\it i.e.}, $\ket{\eta_\pm}$ covers the subspace spanned by $\ket{0}$ \& $\ket{1}$ whereas $\ket{\xi_\pm}$ covers the subspace spanned by $\ket{1}$ \& $\ket{2}$. As a result even if Alice comes together with either Bob or Charlie, it is not possible to perfectly distinguish this set of states. Similar argument holds in case of other bipartitions and consequently $\mathbb{B}_{II}(3,3)$ turns out to be a GNPB of Type-II. Note that this particular basis appears in a recent work for a different purpose \cite{Agrawal19}. There the aim was to construct a genuinely entangled subspace such that all density matrices supported on it are genuinely entangled. Entanglement assisted discrimination protocol of this basis is described in the next section.
 
As already discussed, from a Type-II GNPB no state can be eliminated under OPM when all the parties are separated. However when two party come together the power of state elimination under OPM may increase. This leads us to the further classification of the Type-II GNPBs. 

{\bf Type-II(a):} Such a GNPB is not locally reducible when all the parties are in separate location. But, when two of the parties come together then it is possible to eliminate some states through nontrivial OPM. A careful observation reveals that the GNPB in Proposition \ref{prop2} is an example of such kind (evidently this follows from Remark \ref{remark2} in Appendix \ref{appen-prop8}).  
  
 \begin{figure}[t!]
	\includegraphics[width=0.35\textwidth]{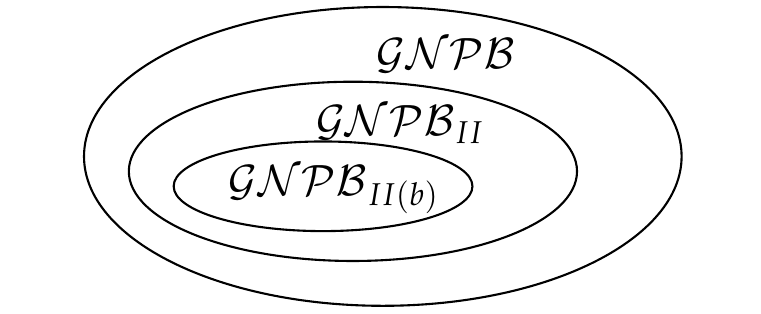}
 	\label{Fig:subset}
 	\caption{The set $\mathcal{GNPB}_{II(b)}$ of all Type-II(b) GNPBs is a proper subset of the set $\mathcal{GNPB}_{II}$ of all Type-II GNPBs which is again a proper subset of the set $\mathcal{GNPB}$ of all GNPBs, {\it i.e.}, $\mathcal{GNPB}_{II(b)}\subset \mathcal{GNPB}_{II}\subset \mathcal{GNPB}$.}
 	\label{subset}
 \end{figure}

{\bf Type-II(b):} Such a GNPB is locally irreducible even if any two of parties come together. Examples of such GNPBs are constructed in \cite{Halder19} for Hilbert spaces $(\mathbb{C}^3)^{\otimes3}$ and $(\mathbb{C}^4)^{\otimes3}$. Here we redraft the $3$-qutrit example. 
\begin{proposition}\label{prop4}
(Halder et al. \cite{Halder19}) The set of states,
\begin{align*}
\rotatebox[origin=c]{0}{$\mathbb{B}_{II(b)}(3,3)\equiv$}
\left\{\!\begin{aligned}
\tri{0}{1}{\eta_\pm},~\tri{1}{\eta_\pm}{0},~\tri{\eta_\pm}{0}{1},\\
\tri{0}{2}{\kappa_\pm},~\tri{2}{\kappa_\pm}{0},~\tri{\kappa_\pm}{0}{2},\\
\tri{1}{2}{\eta_\pm},~\tri{2}{\eta_\pm}{1},~\tri{\eta_\pm}{1}{2},\\
\tri{2}{1}{\kappa_\pm},~\tri{1}{\kappa_\pm}{2},~\tri{\kappa_\pm}{2}{1},\\
\tri{k}{k}{k}~~|~~k\in\{0,1,2\}~~~~
\end{aligned}\right\},	
\end{align*}
in a GNPB of Type-II(b) in $(\mathbb{C}^3)^{\otimes3}$; $\ket{\kappa_\pm}:=(\ket{0}\pm\ket{2})/\sqrt{2}$.
\end{proposition}
Clearly Type-II(b) is the strongest form of GNPB from the perspective of local elimination. The above classification thus introduces a hierarchical relation as depicted in Fig. \ref{subset}.
\section{Entanglement assisted discrimination}\label{sec4}
In this section we study entanglement assisted discrimination protocols for the GNPBs discussed earlier. First we consider the $3$-qutrit GNPBs. Note that, in $(\mathbb{C}^3)^{\otimes3}$ two pairs of $2$-qutrit maximally entangled states, {\it i.e.}, $2\log3$ ebits, distributed between Alice \& Bob and Alice \& Charlie always lead to perfect discrimination for any genuinely nonlocal basis. Therefore any protocol that consumes less than $2\log3$ ebits is nontrivial and resource efficient. Following proposition constitutes such a nontrivial protocol.
\begin{proposition}\label{prop5}
The entanglement resource $\left\lbrace\left(1,\ket{\phi^+(3)}_{\mathcal{AB}}\right);\left(0,\ket{\phi^+}_{\mathcal{BC}}\right);\left(1,\ket{\phi^+}_{\mathcal{CA}}\right)\right\rbrace$ is sufficient for local discrimination of the GNPBs $\mathbb{B}_{II}(3,3)$ and $\mathbb{B}_{II(b)}(3,3)$, where $\ket{\phi^+(3)}:=(\ket{00}+\ket{11}+\ket{22})/\sqrt{3}\in\mathbb{C}^3\otimes\mathbb{C}^3$.  
\end{proposition}
Using $2$-qutrit maximally entangled state $(\ket{00}+\ket{11}+\ket{22})/\sqrt{3}$ Bob first teleports his subsystem to Alice. Entanglement consumed at this step amounts to $\log3$-ebits. After that $1$-ebit entanglement shared between Alice and Charlie suffices for perfect discrimination of these GNPBs (see Appendix \ref{appen-prop5} for detailed protocol). Therefore, in total, $(\log3+1)$-ebits entanglement are consumed in this protocol which is strictly less than the amount consumed in the protocol using teleportation in both arms. However, in this protocol teleportation scheme is used in one arm. We now show that even more efficient protocols are possible to discriminate these GNPBs.      
\begin{proposition}\label{prop6}
The entanglement resource $\left\lbrace\left(1,\ket{\phi^+}_{\mathcal{AB}}\right);\left(0,\ket{\phi^+}_{\mathcal{BC}}\right);\left(1,\ket{\phi^+}_{\mathcal{CA}}\right)\right\rbrace$ sufficiently discriminates the GNPB $\mathbb{B}_{II}(3,3)$ when all the parties are separated.
\end{proposition} 
See Appendix \ref{appen-prop6} for the protocol. Clearly the entanglement consumed in this protocol is strictly less than $(1+\log3)$-ebits. The resource state used in the protocol, {\it i.e.}, the state $\ket{\phi^+}_{\mathcal{AB}}\otimes\ket{\phi^+}_{\mathcal{CA}}$, lives in the Hilbert space $\mathbb{C}^4\otimes\mathbb{C}^2\otimes\mathbb{C}^2$. Naturally the question arises weather a lower dimensional resource from the Hilbert space $\mathbb{C}^2\otimes\mathbb{C}^2\otimes\mathbb{C}^2$ will suffice for perfect discrimination. At this point we observe that a similar protocol like Proposition \ref{prop6} that uses a $3$-qubit GHZ state $\ket{G}$ or a $3$-qubit W-state $\ket{W}:=(\ket{001}+\ket{010}+\ket{100})/\sqrt{3}$ does not lead to perfect discrimination of the basis $\mathbb{B}_{II}(3,3)$. 

Here we want to point out some important observations. For discriminating the NPB of $(\mathbb{C}^3)^{\otimes2}$, Cohen pointed out that in his protocol $2$-qubit maximally entangled state is the necessary resource. If instead of the $2$-qubit maximally entangled resource a partially entangled state $\lambda_0\ket{00}+\lambda_1\ket{11}$, with $\lambda_0\neq\lambda_1$ is provided as resource then at some stage the protocol leads to non orthogonal states and consequently the protocol does not succeed perfectly. Furthermore, he also gave an impression that for any successful protocol $2$-qubit maximally entangled state may be the necessary resource. To the best of our knowledge, however, the assertion is yet to be proven. If it indeed turns out to be the case then the $3$-qubit W-state can not be a sufficient resource for perfect discrimination of the basis $\mathbb{B}_{II}(3,3)$. On the other hand, if there exist no local protocol that simultaneously generates two EPR pairs shared between Alice \& Bob and Alice \& Charlie respectively from a $3$-qubit GHZ state, then it can also not be the sufficient resource for perfect discrimination of $\mathbb{B}_{II}(3,3)$. Therefore, at this point the question remains open whether the resource mentioned in Proposition \ref{prop6} is indeed the necessary resource for perfect discrimination of the set $\mathbb{B}_{II}(3,3)$. 

Using the same resource as of Proposition \ref{prop6} we then proceed to discriminate the set $\mathbb{B}_{II(b)}(3,3)$ following analogous protocol. We find that with this amount of resource though some state can be eliminated, but after a certain stage the protocol can not be further extended up to perfect discrimination. However, if further entanglement resource is provided then a perfect discrimination protocol is possible as stated in the following proposition (detailed protocol provided in Appendix \ref{appen-prop7}).
\begin{proposition}\label{prop7}
The entanglement resource $\left\lbrace\left(1,\ket{\phi^+}_{\mathcal{AB}}\right);\left(\frac{8}{27},\ket{\phi^+}_{\star}\right);\left(1,\ket{\phi^+}_{\mathcal{CA}}\right)\right\rbrace$ is sufficient for perfect local discrimination of the GNPB $\mathbb{B}_{II(b)}(3,3)$, where $\star$ be the one of the pairs from $\{\mathcal{AB},\mathcal{BC},\mathcal{CA}\}$.
\end{proposition} 
The average entanglement used in this protocol is therefore $(1+1+\frac{8}{27})\cong2.296$-ebits which is less than $(1+\log3)\cong2.585$-ebits consumed in Proposition \ref{prop5}. Note that the resource state lives either in the Hilbert space $\mathbb{C}^8\otimes\mathbb{C}^4\otimes\mathbb{C}^2$ (if the addition entanglement in Proposition \ref{prop7} is shared between Alice \& Bob) or in the Hilbert space $\mathbb{C}^4\otimes\mathbb{C}^4\otimes\mathbb{C}^4$ (if the addition entanglement is shared between Bob \& Charlie). At this point the question remains open whether a lower dimensional resource state from $\mathbb{C}^4\otimes\mathbb{C}^2\otimes\mathbb{C}^2$ will result in a successful discrimination protocol for $\mathbb{B}_{II(b)}(3,3)$ or the resource in Proposition \ref{prop7} is necessary. 

Let us now consider entanglement assisted discrimination protocol for the GNPBs in $(\mathbb{C}^4)^{\otimes3}$. As already discussed, for the GNPB $\mathbb{B}_I(4,3)$ in Proposition \ref{prop1} entanglement resource only in one cut is sufficient for perfect discrimination. Of-course in which cut the entangled resource needs to be shared that is determined after the first elimination step under OPM. Furthermore in this case, since the genuine indistinguishability arises due to the presence of $(\mathbb{C}^3)^{\otimes2}$ NPB between Alice \& Bob ({\it i.e.}, the set of states $\left\lbrace \ket{\beta}_{AB}\ket{3}_{C}\right\rbrace$) and between Bob \& Charlie ({\it i.e.} the set of states$\left\lbrace \ket{3}_A\ket{\beta}_{BC}\right\rbrace$), thus Cohen's protocol \cite{Cohen08} assures that a $2$-qubit maximally entangled state shared between $AB$ or $BC$ (decided accordingly after the first elimination step) is sufficient for prefect discrimination even though the local dimension for each party is four.

Consider now the GNPB $\mathbb{B}_{II}(4,3)$ in Proposition \ref{prop2}. Since this one is a GNPB of Type-II, therefore no state can be eliminated under OPM while all the parties are spatially separated and consequently entangled resource across every bipartition is required in this case. Interestingly, here we find that if the three parties share a $3$-qubit GHZ state then they can start the discrimination protocol. However, as stated in the following proposition the perfect discrimination protocol we obtain requires additional EPR pair along with the GHZ resource.   
\begin{proposition}\label{prop8}
	The entanglement resource $\left\lbrace\left(1,\ket{G}_{\mathcal{ABC}}\right);\left(\frac{1}{8},\ket{\phi^+}_{\star}\right)\right\rbrace$ is sufficient for perfect local discrimination of the GNPB $\mathbb{B}_{II}(4,3)$ in Proposition \ref{prop2}, where $\star$ be the one of the pairs from $\{\mathcal{AB},\mathcal{BC},\mathcal{CA}\}$.
\end{proposition}
See Appendix \ref{appen-prop8} for detailed protocol. The above protocol exhibits nontrivial use of multipartite entanglement in local state discrimination protocol. Moreover, we observe that instead of GHZ state if Alice and Bob start the protocol with sharing a EPR state then for perfect discrimination further $\frac{11}{16}$-ebits is required to be shared between Bob and Charlie (see the Remark \ref{remark2} in Appendix \ref{appen-prop8}). This indicates advantage of genuine entangled resource over its bipartite counterpart in state discrimination problem. However, a conclusive proof of this assertion requires establishing the necessary requirement of entangled resources in different such protocols which we leave here as an open question for future research.      
\section{Summary and Open problems}\label{sec5}
The phenomena `strong quantum nonlocality without entanglement' introduced in Ref.~\cite{Halder19} motivates us to look for other techniques to construct new GNPBs. As a consequence, in this work we have classified GNPBs into different categories. Via this classification we have addressed an important question regarding the requirement of multipartite entangled resource state for perfect discrimination of a GNPB. Interestingly, we have found that elimination of certain states from the original set by performing orthogonality-preserving measurements may help to reduce the entanglement consumption for perfect discrimination. We have also presented entanglement assisted local discrimination protocols of several GNPBs. These protocols are resource efficient as they consume less entanglement than a teleportation based protocol. We have also addressed an open problem raised in Ref.~\cite{Halder19}. The authors there left open the possibility of existence of a cheaper resource than that of a teleportation based scheme for perfect discrimination of a strong nonlocal basis. One of our protocols provides affirmative answer to this question. Moreover, we have discrimination protocols for GNPBs with different types and configurations of entangled resources. Interestingly, we find strong indication of genuine entanglement advantage over bipartite entanglement for discrimination of some GNPBs.    

 Our study also raises few important questions. First of all, the question of optimality of the entangled resources used in our discrimination protocols remains open. Clearly, this study will shed light on the optimal resource requirement for the implementation of the separable measurement corresponding to these GNPBs. Furthermore, it is intriguing to study whether the classification of GNPBs induces a hierarchy among the corresponding separable measurements.     

\begin{acknowledgments}
SR acknowledges the support through the VASP programme of S. N. Bose National Center for Basic Sciences. AM acknowledges his visit at S. N. Bose National Center for Basic Sciences. MB acknowledges support through an INSPIRE-faculty position at S. N. Bose National Center for Basic Sciences by the Department of Science and Technology, Government of India.
\end{acknowledgments}

\newpage
\onecolumngrid
\section*{Appendix}
\appendix
\section{Proof of Proposition \ref{prop5}}\label{appen-prop5}
{\bf Discrimination of $\mathbb{B}_{II}(3,3)$:} The GNPB of Proposition \ref{prop3} is given by,
\begin{align}
\rotatebox[origin=c]{0}{$\mathbb{B}_{II}(3,3)\equiv$}
\left\{\!\begin{aligned}
\ket{\psi(\pm,\pm)}_1:=\ket{0}_A\ket{\eta_\pm}_B\ket{\xi_\pm}_C,~\ket{\psi(\pm,\pm)}_2:=\ket{\eta_\pm}_A\ket{2}_B\ket{\xi_\pm}_C,~\\
\ket{\psi(\pm,\pm)}_3:=\ket{2}_A\ket{\xi_\pm}_B\ket{\eta_\pm}_C,~\ket{\psi(\pm,\pm)}_4:=\ket{\eta_\pm}_A\ket{\xi_\pm}_B\ket{0}_C,~\\
\ket{\psi(\pm,\pm)}_5:=\ket{\xi_\pm}_A\ket{0}_B\ket{\eta_\pm}_C,~\ket{\psi(\pm,\pm)}_6:=\ket{\xi_\pm}_A\ket{\eta_\pm}_B\ket{2}_C,~\\
\ket{\phi(k)}:=\ket{k}_A\ket{k}_B\ket{k}_C~~|~~ k\in\{0,1,2\}~~~~~~~~~~~~~~~~
\end{aligned}\right\}.	
\end{align}
Using the entanglement resource $\ket{\phi^+(3)}$ Bob teleports his subsystem to Alice. Thus without loss of generality after this step we can think that they are in same lab and we will use the subindex $\tilde{A}$ for this joint part.  While Alice and Bob are together the states in $\mathbb{B}_{II}(3,3)$ have the following tile structure.
\begin{figure}[h!]
	\includegraphics[width=0.6\textwidth]{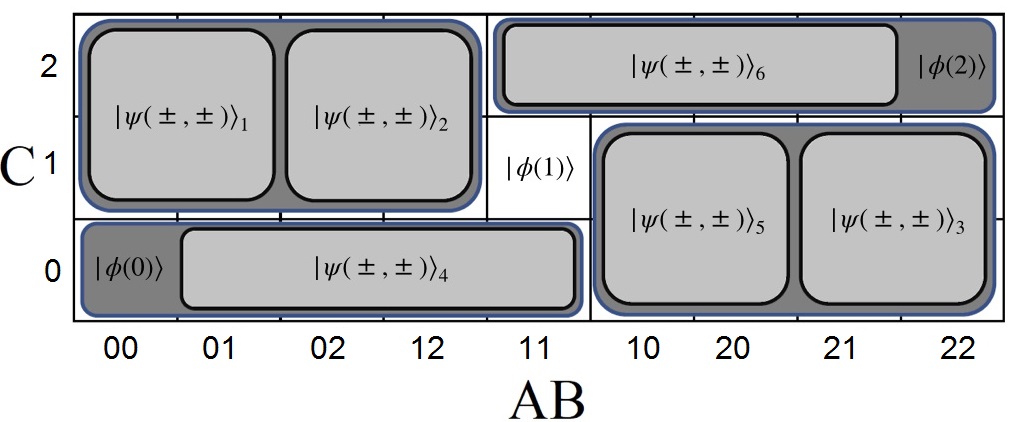}
	\caption{Tile structure of the GNPB $\mathbb{B}_{II}(3,3)$ in AB|C cut. This particular tile structure is similar to that of $\mathbb{C}^3\otimes\mathbb{C}^3$ tile UPB.}
	\label{tile}
\end{figure}

\begin{figure}[h!]
	\includegraphics[width=.8\textwidth]{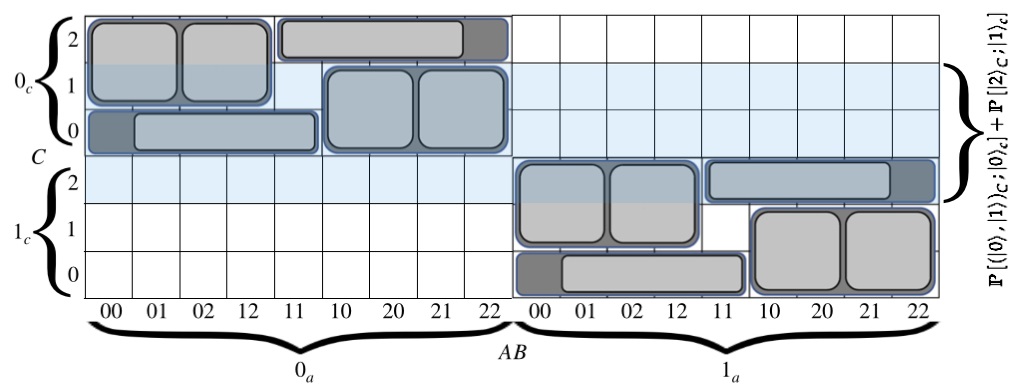}
		\caption{While Alice and Bob are together, the state $\ket{\psi}_{\tilde{A}C}\otimes\ket{\phi^+}_{ac}$ lives in $\mathbb{C}^{18}\otimes\mathbb{C}^6$. Curly brace on right hand side denotes the measurement effect $N:=\mathbb{P}\left[\left(\ket{0},\ket{1}\right)_C;\ket{0}_{c}\right]+\mathbb{P}\left[\ket{2}_C;\ket{1}_{c}\right]$ in Step-1.}
		\label{tile1}
\end{figure}

For discriminating the state Charlie shares $\ket{\phi^+}$ with $\tilde{A}$. Therefore the initial state is 
\begin{equation}
\ket{\psi}_{\tilde{A}C}\otimes\ket{\phi^+}_{ac},
\end{equation} 
where $\ket{\psi}_{\tilde{A}C}$ is one of the states from $\mathbb{B}_{II}(3,3)$ and this allows representation as in Fig. \ref{tile1} (see \cite{Cohen08} for details of this representation).

For short hand notation we will denote $\ket{ij}\mapsto \ket{{\bf 3i+j}}$. Now the discrimination protocol proceeds as follows.

{\bf Step-1:} Charlie performs the measurement
$$\mathcal{N}\equiv\left\lbrace N:=\mathbb{P}\left[\left(\ket{0},\ket{1}\right)_C;\ket{0}_{c}\right]+\mathbb{P}\left[\ket{2}_C;\ket{1}_{c}\right],\overline{N}:=\mathbb{I}-N\right\rbrace,$$
where, $\mathbb{P}\left[\left(\ket{i},\ket{j}\right)_{\$};\left(\ket{k},\ket{l}\right)_{\#}\right]:=\left( \ket{i}\bra{i}+\ket{j}\bra{j}\right)_{\$}\otimes\left( \ket{k}\bra{k}+\ket{l}\bra{l}\right)_{\#}$, and  this definition is applicable for all the protocols. Suppose outcome corresponding to $N$ clicks.

{\bf Step-2:} Alice performs the measurement
\begin{align}
\rotatebox[origin=c]{0}{$\mathcal{K}\equiv$}
\left\{\!\begin{aligned}
K_1&:=\mathbb{P}\left[\left(\ket{\mathbf{3}},\ket{\mathbf{6}},\ket{\mathbf{7}},\ket{\mathbf{8}}\right)_{\tilde{A}};\ket{0}_{a}\right],\\
K_2&:=\mathbb{P}\left[\left(\ket{\mathbf{3}},\ket{\mathbf{4}},\ket{\mathbf{6}},\ket{\mathbf{7}},\ket{\mathbf{8}}\right)_{\tilde{A}};\ket{1}_{a}\right],\\
K_3&:=\mathbb{I}-K_1-K_2.~~~~~~~~~~
\end{aligned}\right\}.	
\end{align}
If $K_1$ clicks the given state is one of $\{\ket{\psi(\pm,\pm)}_3,\ket{\psi(\pm,\pm)}_5\}$ and this set of states are perfectly LOCC distinguishable. If $K_1$ clicks the state is one of $\{\ket{\psi(\pm,\pm)}_6,\ket{\phi(2)}\}$, again a LOCC distinguishable set. Otherwise it is one of the remaining $14$ states.

{\bf Step-3:} Charlie performs the measurement
$\mathcal{N}^\prime\equiv\left\lbrace N^\prime:=\mathbb{P}\left[\ket{0}_C;\mathbb{I}_{c}\right],\overline{N}^\prime:=\mathbb{I}-N^\prime\right\rbrace$. If $N^\prime$ clicks the state is one of $\{\ket{\psi(\pm,\pm)}_4,\ket{\phi(0)}\}$ (LOCC distinguishable set), else it is one of remaining $9$ states.

{\bf Step-4:} Alice performs the measurement $\mathcal{K}^\prime\equiv\left\lbrace K^\prime:=\mathbb{P}\left[\ket{\mathbf{4}}_{\tilde{A}};\mathbb{I}_{a}\right],\overline{K}^\prime:=\mathbb{I}-K^\prime\right\rbrace$. If $K^\prime$ clicks the state is $\ket{\phi(1)}$, otherwise it is one of $\{\ket{\psi(\pm,\pm)}_1,\ket{\psi(\pm,\pm)}_2\}$, a LOCC distinguishable set. If in {\bf Step-1} $\overline{N}$ clicks then also a similar protocol follows.

{\bf Discrimination of $\mathbb{B}_{II(b)}(3,3)$:} The GNPB of Proposition \ref{prop3} is given by,
\begin{align}
\rotatebox[origin=c]{0}{$\mathbb{B}_{II(b)}(3,3)\equiv$}
\left\{\!\begin{aligned}
\ket{\alpha(\pm)}_1:=\toph{0}{1}{\eta_\pm},~~\ket{\alpha(\pm)}_2:=\toph{0}{2}{\kappa_\pm},\\
\ket{\alpha(\pm)}_3:=\toph{1}{2}{\eta_\pm},~~\ket{\alpha(\pm)}_4:=\toph{2}{1}{\kappa_\pm},\\
\ket{\beta(\pm)}_1:=\toph{1}{\eta_\pm}{0},~~\ket{\beta(\pm)}_2:=\toph{2}{\kappa_\pm}{0},\\
\ket{\beta(\pm)}_3:=\toph{2}{\eta_\pm}{1},~~\ket{\beta(\pm)}_4:=\toph{1}{\kappa_\pm}{2},\\
\ket{\gamma(\pm)}_1:=\toph{\eta_\pm}{0}{1},~~\ket{\gamma(\pm)}_2:=\toph{\kappa_\pm}{0}{2},\\
\ket{\gamma(\pm)}_3:=\toph{\eta_\pm}{1}{2},~~\ket{\gamma(\pm)}_4:=\toph{\kappa_\pm}{2}{1},\\
\ket{\phi(k)}:=\toph{k}{k}{k}~~|~~k\in\{0,1,2\}~.~~~~~~~~~~
\end{aligned}\right\}.	
\end{align}
While Alice and Bob are together (after Bob teleports his part to Alice using $\log3$-ebits) the states in $\mathbb{B}_{II(b)}(3,3)$ have the following tile structure.
\begin{figure}[h!]
	\includegraphics[width=.6\textwidth]{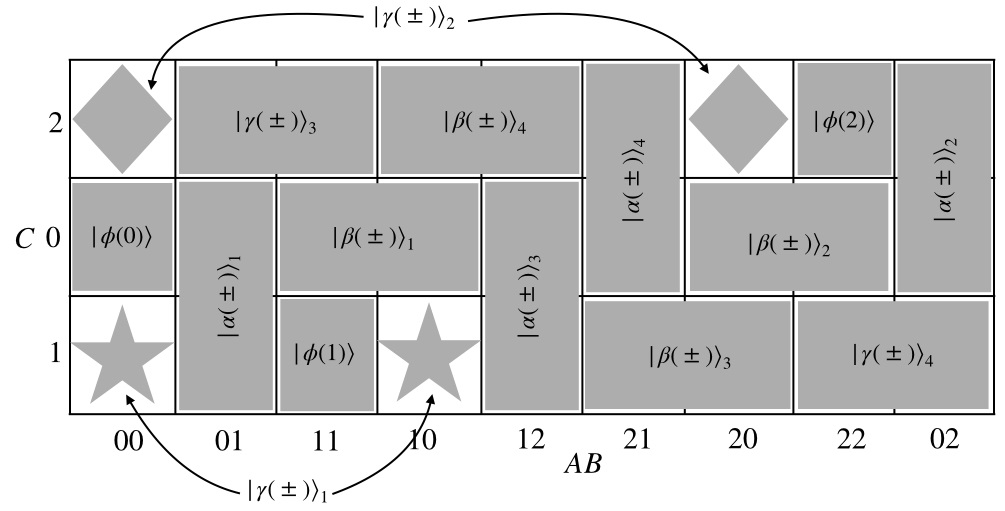}
	\caption{Tile structure of the GNPB $\mathbb{B}_{II(b)}(3,3)$ in AB|C cut. The star (diamond) shaped tiles contain the states $\ket{\gamma(\pm)}_1~(\ket{\gamma(\pm)}_2)$.}
	\label{tile2}
\end{figure}

{\bf Step-1:} Charlie performs the measurement
$$\mathcal{N}\equiv\left\lbrace N:=\mathbb{P}\left[\left(\ket{0},\ket{1}\right)_C;\ket{0}_{c}\right]+\mathbb{P}\left[\ket{2}_C;\ket{1}_{c}\right],\overline{N}:=\mathbb{I}-N\right\rbrace.$$
Suppose $N$ clicks.

{\bf Step-2:} Alice's measurement and the states corresponding to different outcomes are shown.
\begin{align}
\rotatebox[origin=c]{0}{$\mathcal{K}\equiv$}
\left\{\!\begin{aligned}
K_1&:=\mathbb{P}\left[\left(\ket{\mathbf{0}},\ket{\mathbf{3}},\ket{\mathbf{4}}\right)_{\tilde{A}};\ket{0}_{a}\right]\Rightarrow\{\ket{\beta(\pm)}_1,\ket{\gamma(\pm)}_1,\ket{\phi(0)},\ket{\phi(1)}\}\\
K_2&:=\mathbb{P}\left[\ket{\mathbf{1}}_{\tilde{A}};\ket{0}_{a}\right]\Rightarrow\{\ket{\alpha(\pm)}_1\},~~~~~~~~~~~~~~
K_3:=\mathbb{P}\left[\ket{\mathbf{5}}_{\tilde{A}};\ket{0}_{a}\right]\Rightarrow\{\ket{\alpha(\pm)}_3\},\\
K_4&:=\mathbb{P}\left[\left(\ket{\mathbf{0}},\ket{\mathbf{6}}\right)_{\tilde{A}};\ket{1}_{a}\right]\Rightarrow\{\ket{\gamma(\pm)}_2\},~~~~
K_5:=\mathbb{P}\left[\left(\ket{\mathbf{1}},\ket{\mathbf{4}}\right)_{\tilde{A}};\ket{1}_{a}\right]\Rightarrow\{\ket{\gamma(\pm)}_3\},\\
K_6&:=\mathbb{P}\left[\left(\ket{\mathbf{3}},\ket{\mathbf{5}}\right)_{\tilde{A}};\ket{1}_{a}\right]\Rightarrow\{\ket{\beta(\pm)}_4\},~~~~
K_7:=\mathbb{P}\left[\ket{\mathbf{8}}_{\tilde{A}};\ket{1}_{a}\right]\Rightarrow\{\ket{\phi(2)}\},\\
K_8&:=\mathbb{I}-\sum_{i=1}^7K_i\Rightarrow\{\ket{\alpha(\pm)}_2,\ket{\alpha(\pm)}_4,\ket{\beta(\pm)}_2,\ket{\beta(\pm)}_3,\ket{\gamma(\pm)}_4\}.
\end{aligned}\right\}.	
\end{align}

{\bf Step-3:} If $K_1$ clicks, Charlie performs the measurement $\mathcal{N}^\prime\equiv\left\lbrace N^\prime:=\mathbb{P}\left[\ket{0}_C;\mathbb{I}_{c}\right],\overline{N}^\prime:=\mathbb{I}-N^\prime\right\rbrace$. If $N^\prime$ clicks then Alice performs the measurement $\mathcal{K}^\prime\equiv\left\lbrace K^\prime:=\mathbb{P}\left[\ket{\mathbf{0}}_{\tilde{A}};\mathbb{I}_{a}\right],\overline{K}^\prime:=\mathbb{I}-K^\prime\right\rbrace$, else she performs $\mathcal{K}^\prime\equiv\left\lbrace K^\prime:=\mathbb{P}\left[\ket{\mathbf{4}}_{\tilde{A}};\mathbb{I}_{a}\right],\overline{K}^\prime:=\mathbb{I}-K^\prime\right\rbrace$. The states corresponding to the outcomes are listed below:
\begin{align}
\left\{\!\begin{aligned}
N^\prime~\&~K^\prime&\Rightarrow\ket{\phi(0)},~~
N^\prime~\&~\overline{K}^\prime\Rightarrow\{\ket{\beta(\pm)}_1\},\\
\overline{N}^\prime~\&~ K^\prime&\Rightarrow\ket{\phi(1)},~~
\overline{N}^\prime~\&~\overline{K}^\prime\Rightarrow\{\ket{\gamma(\pm)}_1\}.
\end{aligned}\right\}.	
\end{align}

If $K_8$ clicks, Charlie performs the measurement $\mathcal{N}^\prime\equiv\left\lbrace N^\prime:=\mathbb{P}\left[\ket{1}_C;\mathbb{I}_{c}\right],\overline{N}^\prime:=\mathbb{I}-N^\prime\right\rbrace$. If $N^\prime$ clicks then Alice performs the measurement $\mathcal{K}^\prime\equiv\left\lbrace K^\prime:=\mathbb{P}\left[\left(\ket{\mathbf{6}},\ket{\mathbf{7}}\right) _{\tilde{A}};\mathbb{I}_{a}\right],\overline{K}^\prime:=\mathbb{I}-K^\prime\right\rbrace$, else she performs $\mathcal{K}^\prime\equiv\left\lbrace K_1^\prime:=\mathbb{P}\left[\ket{\mathbf{7}}_{\tilde{A}};\mathbb{I}_{a}\right],K_2^\prime:=\mathbb{P}\left[\ket{\mathbf{2}}_{\tilde{A}};\mathbb{I}_{a}\right] ,K_3^\prime:=\mathbb{I}-K_1^\prime-K_2^\prime\right\rbrace $. The states corresponding to the outcomes are listed below:
\begin{align}
\left\{\!\begin{aligned}
N^\prime~\&~K^\prime\Rightarrow\{\ket{\beta(\pm)}_3\},~~
N^\prime~\&~\overline{K}^\prime\Rightarrow\{\ket{\gamma(\pm)}_4\},\\
\overline{N}^\prime~\&~ K_1^\prime\Rightarrow\{\ket{\alpha(\pm)}_4\},~~
\overline{N}^\prime~\&~K_2^\prime\Rightarrow\{\ket{\alpha(\pm)}_2\},\\
\overline{N}^\prime~\&~K_3^\prime\Rightarrow\{\ket{\beta(\pm)}_2\}.~~~~~~~~~~~~~~~~
\end{aligned}\right\}.	
\end{align}
In {\bf Step-1} if $\overline{N}$ clicks instead of $N$ then also a similar protocol follows.

\section{Proof of Proposition \ref{prop6}}\label{appen-prop6}
We need to discriminate the basis $\mathbb{B}_{II}(3,3)$. Let the EPR state between Alice \& Bob be denoted as $\ket{\phi^+}_{a_1b_1}$ and that shared between Alice \& Charlie be denoted as $\ket{\phi^+}_{a_2c_1}$. Therefore the initial shared states among them is,
\begin{equation}
\ket{\psi}_{ABC}\otimes\ket{\phi^+}_{a_1b_1}\otimes\ket{\phi^+}_{a_2c_1},
\end{equation}
where $\ket{\psi}_{ABC}$ is one of the state from the set $\mathbb{B}_{II}(3,3)$.

{\bf Step-1:} Bob performs a measurement
$$\mathcal{M}\equiv\left\lbrace M:=\mathbb{P}\left[\left(\ket{0},\ket{1}\right)_B;\ket{0}_{b_1}\right]+\mathbb{P}\left[\ket{2}_B;\ket{1}_{b_1}\right],\overline{M}:=\mathbb{I}-M\right\rbrace,$$
and Charlie performs measurement,
$$\mathcal{N}\equiv\left\lbrace N:=\mathbb{P}\left[\left(\ket{1},\ket{2}\right)_C;\ket{0}_{c_1}\right]+\mathbb{P}\left[\ket{0}_C;\ket{1}_{c_1}\right],\overline{N}:=\mathbb{I}-N\right\rbrace.$$
Suppose the outcomes corresponding to $M$ and $N$ click. The resulting post measurement state is therefore,
\begin{align}\label{5s1}
\left\{\!\begin{aligned}
\ket{\psi(\pm,\pm)}_1&\longrightarrow\ket{0}_A\ket{\eta_\pm}_B\ket{\xi_\pm}_C\ket{00}_{a_1b_1}\ket{00}_{a_2c_1},\\
\ket{\psi(\pm,\pm)}_2&\longrightarrow\ket{\eta_\pm}_A\ket{2}_B\ket{\xi_\pm}_C\ket{11}_{a_1b_1}\ket{00}_{a_2c_1},\\
\ket{\psi(\pm,\pm)}_3&\longrightarrow\ket{2}_A\left(\ket{1}_B\ket{00}_{a_1b_1}\pm\ket{2}_B\ket{11}_{a_1b_1} \right)\left(\ket{0}_C\ket{11}_{a_2c_1}\pm\ket{1}_C\ket{00}_{a_2c_1}\right),\\
\ket{\psi(\pm,\pm)}_4&\longrightarrow\ket{\eta_\pm}_A\left(\ket{1}_B\ket{00}_{a_1b_1}\pm\ket{2}_B\ket{11}_{a_1b_1} \right)\ket{0}_C\ket{11}_{a_2c_1},\\
\ket{\psi(\pm,\pm)}_5&\longrightarrow\ket{\xi_\pm}_A\ket{0}_B\ket{00}_{a_1b_1}\left(\ket{0}_C\ket{11}_{a_2c_1}\pm\ket{1}_C\ket{00}_{a_2c_1}\right),\\
\ket{\psi(\pm,\pm)}_6&\longrightarrow\ket{\xi_\pm}_A\ket{\eta_\pm}_B\ket{2}_C\ket{00}_{a_1b_1}\ket{00}_{a_2c_1},\\
\ket{\phi(0)}&\longrightarrow\toph{0}{0}{0}\ket{00}_{a_1b_1}\ket{11}_{a_2c_1},\\
\ket{\phi(1)}&\longrightarrow\toph{1}{1}{1}\ket{00}_{a_1b_1}\ket{00}_{a_2c_1},\\
\ket{\phi(2)}&\longrightarrow\toph{2}{2}{2}\ket{11}_{a_1b_1}\ket{00}_{a_2c_1}.
\end{aligned}\right\}.	
\end{align}

{\bf Step-2:} Alice performs the measurement,
$$\mathcal{K}\equiv\left\lbrace K_1:=\mathbb{P}\left[\left(\ket{0},\ket{1}\right)_A;\ket{1}_{a_1};\ket{0}_{a_2}\right],K_2:=\mathbb{P}\left[\ket{0}_A;\ket{0}_{a_1};\ket{0}_{a_2}\right],K_3:=\mathbb{I}-K_1-K_2\right\rbrace.$$
If $K_1$ clicks, the given state is from the set $\{\ket{\psi(\pm,\pm)}_2\}$ which is LOCC distinguishable. If, $K_2$ clicks, the state is one of $\{\ket{\psi(\pm,\pm)}_1\}$ (LOCC distinguishable set). Else, the state is one of the remaining $19$ states. 

{\bf Step-3:} Charlie performs the measurement, $\mathcal{N}^\prime\equiv\left\lbrace N^\prime:=\mathbb{P}\left[\ket{2}_C;\mathbb{I}_{c_1}\right],\overline{N}^\prime:=\mathbb{I}-N^\prime\right\rbrace.$ If $N^\prime$ clicks, the state is one of $\{\ket{\psi(\pm,\pm)}_6,\ket{\phi(2)}\}$ which is perfectly LOCC distinguishable. 

{\bf Step-4:} Bob performs the measurement, $\mathcal{M}^\prime\equiv\left\lbrace M^\prime:=\mathbb{P}\left[\ket{0}_B;\mathbb{I}_{b_1}\right],\overline{M}^\prime:=\mathbb{I}-M^\prime\right\rbrace.$ If $M^\prime$ clicks, the state is one of $\{\ket{\psi(\pm,\pm)}_5,\ket{\phi(0)}\}$ which is again perfectly LOCC distinguishable.

{\bf Step-5:} Alice performs the measurement, $\mathcal{K}^\prime\equiv\left\lbrace K^\prime:=\mathbb{P}\left[\ket{2}_A;\mathbb{I}_{a_1};\mathbb{I}_{a_2}\right],\overline{K}^\prime:=\mathbb{I}-K^\prime\right\rbrace.$ If $K^\prime$ clicks, the state is one of $\{\ket{\psi(\pm,\pm)}_3\}$ (LOCC distinguishable set) else it is from the remaining set of LOCC distinguishable states $\{\ket{\psi(\pm,\pm)}_4,\ket{\phi(1)}\}$.

After the {\bf Step-1}, only one case (corresponding to the outcomes $M$ and $N$) is discussed. For all other cases a similar protocol follows.

\section{Proof of Proposition \ref{prop7}}\label{appen-prop7}
We need to discriminate the set $\mathbb{B}_{II(b)}(3,3)$.
EPR state shared between Alice \& Bob be denoted as $\ket{\phi^+}_{a_1b_1}$, shared between Alice \& Charlie as $\ket{\phi^+}_{a_2c_1}$, and between Bob \& Charlie as $\ket{\phi^+}_{b_2c_2}$. Therefore the initial shared states among them is,
\begin{equation}
\ket{\psi}_{ABC}\otimes\ket{\phi^+}_{a_1b_1}\otimes\ket{\phi^+}_{a_2c_1}\otimes\ket{\phi^+}_{b_2c_2},
\end{equation}
where $\ket{\psi}_{ABC}$ is one of the state from the set $\mathbb{B}_{II(b)}(3,3)$ which they want to identify by LOCC.

{\bf Step-1:} Bob performs a measurement 
$$\mathcal{M}\equiv\left\lbrace M:=\mathbb{P}\left[\left(\ket{0},\ket{1}\right)_B;\ket{0}_{b_1}\right]+\mathbb{P}\left[\ket{2}_B;\ket{1}_{b_1}\right] ,~\overline{M}:=\mathbb{I}-M\right\rbrace.$$ Charlie performs a measurement 
$$\mathcal{N}\equiv\left\lbrace N:=\mathbb{P}\left[\left(\ket{0},\ket{1}\right)_C;\ket{0}_{c_1}\right]+\mathbb{P}\left[\ket{2}_C;\ket{1}_{c_1}\right] ,~\overline{N}:=\mathbb{I}-N\right\rbrace,$$
Suppose that the outcomes corresponding to $M$ and $N$ click. The resulting post measurement state is therefore,

\begin{align}\label{6s1}
\left\{\!\begin{aligned}
\ket{\alpha(\pm)}_1&\longrightarrow\ket{0}_A\ket{1}_B\ket{\eta_\pm}_C\ket{00}_{a_1b_1}\ket{00}_{a_2c_1}\ket{\phi^+}_{b_2c_2},\\
\ket{\alpha(\pm)}_2&\longrightarrow\ket{0}_A\ket{2}_B\ket{11}_{a_1b_1}\left(\ket{0}_C\ket{00}_{a_2c_1}\pm\ket{2}_C\ket{11}_{a_2c_1}\right) \ket{\phi^+}_{b_2c_2},\\
\ket{\alpha(\pm)}_3&\longrightarrow\ket{1}_A\ket{2}_B\ket{\eta_\pm}_C\ket{11}_{a_1b_1}\ket{00}_{a_2c_1}\ket{\phi^+}_{b_2c_2},\\
\ket{\alpha(\pm)}_4&\longrightarrow\ket{2}_A\ket{1}_B\ket{00}_{a_1b_1}\left(\ket{0}_C\ket{00}_{a_2c_1}\pm\ket{2}_C\ket{11}_{a_2c_1}\right)\ket{\phi^+}_{b_2c_2},\\
\ket{\beta(\pm)}_1&\longrightarrow\toph{1}{\eta_\pm}{0}\ket{00}_{a_1b_1}\ket{00}_{a_2c_1}\ket{\phi^+}_{b_2c_2},\\
\ket{\beta(\pm)}_2&\longrightarrow\ket{2}_A\left(\ket{0}_B\ket{00}_{a_1b_1}\pm\ket{2}_B\ket{11}_{a_1b_1}\right) \ket{0}_C\ket{00}_{a_2c_1}\ket{\phi^+}_{b_2c_2},\\
\ket{\beta(\pm)}_3&\longrightarrow\toph{2}{\eta_\pm}{1}\ket{00}_{a_1b_1}\ket{00}_{a_2c_1}\ket{\phi^+}_{b_2c_2},\\
\ket{\beta(\pm)}_4&\longrightarrow\ket{1}_A\left(\ket{0}_B\ket{00}_{a_1b_1}\pm\ket{2}_B\ket{11}_{a_1b_1}\right) \ket{2}_C\ket{11}_{a_2c_1}\ket{\phi^+}_{b_2c_2},\\
\ket{\gamma(\pm)}_1&\longrightarrow\toph{\eta_\pm}{0}{1}\ket{00}_{a_1b_1}\ket{00}_{a_2c_1}\ket{\phi^+}_{b_2c_2},\\
\ket{\gamma(\pm)}_2&\longrightarrow\toph{\kappa_\pm}{0}{2}\ket{00}_{a_1b_1}\ket{11}_{a_2c_1}\ket{\phi^+}_{b_2c_2},\\
\ket{\gamma(\pm)}_3&\longrightarrow\toph{\eta_\pm}{1}{2}\ket{00}_{a_1b_1}\ket{11}_{a_2c_1}\ket{\phi^+}_{b_2c_2},\\
\ket{\gamma(\pm)}_4&\longrightarrow\toph{\kappa_\pm}{2}{1}\ket{11}_{a_1b_1}\ket{00}_{a_2c_1}\ket{\phi^+}_{b_2c_2},\\
\ket{\phi(0)}&\longrightarrow\toph{0}{0}{0}\ket{00}_{a_1b_1}\ket{00}_{a_2c_1}\ket{\phi^+}_{b_2c_2},\\
\ket{\phi(1)}&\longrightarrow\toph{1}{1}{1}\ket{00}_{a_1b_1}\ket{00}_{a_2c_1}\ket{\phi^+}_{b_2c_2},\\
\ket{\phi(2)}&\longrightarrow\toph{2}{2}{2}\ket{11}_{a_1b_1}\ket{11}_{a_2c_1}\ket{\phi^+}_{b_2c_2}.
\end{aligned}\right\}.	
\end{align}
{\bf Step-2:} Alice performs the measurement 
\begin{align}
\rotatebox[origin=c]{0}{$\mathcal{K}\equiv$}
\left\{\!\begin{aligned}
K_1&:=\mathbb{P}\left[\ket{1}_A;\ket{1}_{a_1};\ket{0}_{a_2}\right],~~~~~~~~~~~~~
K_2:=\mathbb{P}\left[\ket{2}_A;\ket{1}_{a_1};\ket{1}_{a_2}\right],\\
K_3&:=\mathbb{P}\left[\left( \ket{0}_A,\ket{1}_A\right) ;\ket{0}_{a_1};\ket{0}_{a_2}\right],~~
K_4:=\mathbb{I}-K_1-K_2-K_3.
\end{aligned}\right\}.	
\end{align}
If $K_1$ clicks the given state $\ket{\psi}_{ABC}$ is one of $\{\ket{\alpha(\pm)}_3\}$ which are LOCC discriminable; if $K_2$ clicks the state is $\ket{\phi(2)}$; if $K_3$ clicks the state is one of the following set of states,
\begin{align}\label{7s7}
\left\{\!\begin{aligned}
\ket{\alpha(\pm)}_1&\rightarrow\toph{0}{1}{\eta_\pm}\ket{00}_{a_1b_1}\ket{00}_{a_2c_1}\ket{\phi^+}_{b_2c_2},\\
\ket{\beta(\pm)}_1&\rightarrow\toph{1}{\eta_\pm}{0}\ket{00}_{a_1b_1}\ket{00}_{a_2c_1}\ket{\phi^+}_{b_2c_2},\\
\ket{\gamma(\pm)}_1&\rightarrow\toph{\eta_\pm}{0}{1}\ket{00}_{a_1b_1}\ket{00}_{a_2c_1}\ket{\phi^+}_{b_2c_2},\\
\ket{\phi(0)}&\rightarrow\toph{0}{0}{0}\ket{00}_{a_1b_1}\ket{00}_{a_2c_1}\ket{\phi^+}_{b_2c_2},\\
\ket{\phi(1)}&\rightarrow\toph{1}{1}{1}\ket{00}_{a_1b_1}\ket{00}_{a_2c_1}\ket{\phi^+}_{b_2c_2}.
\end{aligned}\right\}.	
\end{align}
Since all states of the ancillary systems $a_1b_1$ and $a_2c_1$ are identical in all the case so they provide no further advantage in discrimination and therefore redundant. Detaching theses ancillary systems along with $\ket{\phi^+}_{b_2c_2}$ the remaining states are the complete product bases corresponding to the shift UPB of $\mathbb{C}^2\otimes\mathbb{C}^2\otimes{C}^2$ \cite{Bennett99-1}. These states cannot be further discriminated under LOCC. However, the additional resource state $\ket{\phi^+}_{b_2c_2}$ make it possible to discriminate the above states perfectly.

If $K_4$ clicks the given state is one of the remaining $27-(2+1+8)=16$ states. For these states discriminating protocol goes as follows.

{\bf Step-3:} Charlie performs $\mathcal{N}^\prime\equiv\left\lbrace N^\prime:=\mathbb{P}\left[\ket{1}_C;\mathbb{I}_{c_1}\right],~\overline{N}^\prime:=\mathbb{I}-N^\prime\right\rbrace$. If $N^\prime$ clicks the state is given from $\{\ket{\beta(\pm)}_3,\ket{\gamma(\pm)}_4\}$; else it is one of the remaining $12$ states.  

{\bf Step-4:} Bob performs $\mathcal{M}^\prime\equiv\left\lbrace M^\prime:=\mathbb{P}\left[\ket{1}_B;\mathbb{I}_{b_1}\right],~\overline{M}^\prime:=\mathbb{I}-M^\prime\right\rbrace$. If $M^\prime$ clicks the state is given from $\{\ket{\alpha(\pm)}_4,\ket{\gamma(\pm)}_3\}$; else it is one of the remaining $8$ states.

{\bf Step-5:} Alice performs the measurement
\begin{align}
\rotatebox[origin=c]{0}{$\mathcal{K}^\prime\equiv$}
\left\{\!\begin{aligned}
K_1^\prime&:=\mathbb{P}\left[\ket{0}_A;\ket{1}_{a_1};\mathbb{I}_{a_2}\right],~~~
K_2^\prime:=\mathbb{P}\left[\ket{2}_A;\mathbb{I}_{a_1};\ket{0}_{a_2}\right],\\
K_3^\prime&:=\mathbb{P}\left[\ket{1}_A;\mathbb{I}_{a_1};\mathbb{I}_{a_2}\right],~~~~~~~~
K_4^\prime:=\mathbb{P}\left[\left(\ket{0}_A,\ket{2}_A\right) ;\ket{0}_{a_1};\ket{1}_{a_2}\right]
\end{aligned}\right\}.	
\end{align}
If $K_1^\prime$ clicks the state is one of $\{\ket{\alpha(\pm)}_2\}$, if $K_2^\prime$ clicks the state is one of $\{\ket{\beta(\pm)}_2\}$, if $K_3^\prime$ clicks the state is one of $\{\ket{\beta(\pm)}_4\}$, else the state is one of $\{\ket{\gamma(\pm)}_2\}$.

Since the given state is chosen randomly from the set $\mathbb{B}_{II(b)}(3,3)$, therefore the average entanglement consumption in the above protocol is $(1+1+\frac{8}{27})$-ebits. Here, remember that after the {\bf Step-1}, only one outcome is discussed. Other outcomes are also equally likely and hence the entanglement consumption is actually the average.

\section{Proof of Proposition \ref{prop8}}\label{appen-prop8}
The set of states needs to be discriminated is given by,
\begin{align}\label{c1}
\rotatebox[origin=c]{0}{$\mathbb{B}_{II}(4,3):=$}
\left\{\!\begin{aligned}
\ket{3}\ket{\beta},~\ket{\beta}\ket{3},~\tri{0}{3}{\chi_\pm},~\tri{2}{\chi_\pm}{2},~\tri{\chi_\pm}{3}{1},~~~~~~~~~~~~~\\
|3\rangle|0\rangle|3\rangle,~|3\rangle|1\rangle|3\rangle,~|3\rangle|2\rangle|3\rangle,~|3\rangle|3\rangle|0\rangle,~|3\rangle|3\rangle|2\rangle,~
|3\rangle|3\rangle|3\rangle,~|2\rangle|0\rangle|0\rangle,\\
|2\rangle|0\rangle|1\rangle,~|2\rangle|0\rangle|2\rangle,~|2\rangle|1\rangle|0\rangle,~|2\rangle|1\rangle|1\rangle,~|2\rangle|1\rangle|2\rangle,~
|2\rangle|2\rangle|0\rangle,~|2\rangle|2\rangle|1\rangle,\\
|2\rangle|3\rangle|0\rangle,~|2\rangle|3\rangle|3\rangle,~|0\rangle|0\rangle|0\rangle,~|0\rangle|0\rangle|1\rangle,~|0\rangle|0\rangle|2\rangle,~
|0\rangle|1\rangle|0\rangle,~|0\rangle|1\rangle|1\rangle,\\
|0\rangle|1\rangle|2\rangle,~|0\rangle|2\rangle|0\rangle,~|0\rangle|2\rangle|1\rangle,~|0\rangle|2\rangle|2\rangle,~|0\rangle|3\rangle|0\rangle,~
|0\rangle|3\rangle|1\rangle,~|1\rangle|0\rangle|0\rangle,\\
|1\rangle|0\rangle|1\rangle,~|1\rangle|0\rangle|2\rangle,~|1\rangle|1\rangle|0\rangle,~|1\rangle|1\rangle|1\rangle,~|1\rangle|1\rangle|2\rangle,~
|1\rangle|2\rangle|0\rangle,~|1\rangle|2\rangle|1\rangle,\\
|1\rangle|2\rangle|2\rangle,~|1\rangle|3\rangle|0\rangle,~|1\rangle|3\rangle|1\rangle,~|1\rangle|3\rangle|2\rangle,~|1\rangle|3\rangle|3\rangle.~~~~~~~~~~~~~~~
\end{aligned}\right\},	
\end{align} 
where $\ket{\beta}\in\mathcal{B}\equiv\left\lbrace\bi{0}{\eta_\pm},~\bi{\eta_\pm}{2},~\bi{2}{\xi_\pm},~\bi{\xi_\pm}{0},~\bi{1}{1}\right\rbrace$. Suppose that they share the resource state $(\ket{000}_{abc}+\ket{111}_{abc})/\sqrt{2}$ among them.

{\bf Step-1:} Bob performs the measurement 
$$\mathcal{M}\equiv\left\lbrace M:=\mathbb{P}\left[\left(\ket{0},\ket{1}\right)_B;\ket{0}_{b}\right]+\mathbb{P}\left[\left(\ket{2},\ket{3}\right)_B;\ket{1}_{b}\right] ,~\overline{M}:=\mathbb{I}-M\right\rbrace.$$
Suppose $M$ clicks. The set of states tagged only with $\ket{000}_{abc}$, only with $\ket{111}_{abc}$, or in entangled form of these tags are listed below:
\begin{align}\label{c11}
\rotatebox[origin=c]{0}{$\ket{000}_{abc}\Rightarrow$}
\left\{\!\begin{aligned}
\ket{3}\bi{0}{\eta_\pm},~\ket{3}\bi{\eta_\pm}{2},~\ket{3}\bi{1}{1},~\bi{0}{\eta_\pm}\ket{3},~\bi{\xi_\pm}{0}\ket{3},~\bi{1}{1}\ket{3},\\
|3\rangle|0\rangle|3\rangle,~|3\rangle|1\rangle|3\rangle,~|2\rangle|0\rangle|0\rangle,~|2\rangle|0\rangle|1\rangle,~|2\rangle|0\rangle|2\rangle,
~|2\rangle|1\rangle|0\rangle,~|2\rangle|1\rangle|1\rangle,~~~~\\
|2\rangle|1\rangle|2\rangle,~|0\rangle|0\rangle|0\rangle,~|0\rangle|0\rangle|1\rangle,~|0\rangle|0\rangle|2\rangle,~|0\rangle|1\rangle|0\rangle,
~|0\rangle|1\rangle|1\rangle,~|0\rangle|1\rangle|2\rangle,~~~~\\
|1\rangle|0\rangle|0\rangle,~|1\rangle|0\rangle|1\rangle,~|1\rangle|0\rangle|2\rangle,~|1\rangle|1\rangle|0\rangle,~
|1\rangle|1\rangle|1\rangle,~|1\rangle|1\rangle|2\rangle~~~~~~~~~~~~
\end{aligned}\right\},\\\nonumber\\
\rotatebox[origin=c]{0}{$\ket{111}_{abc}\Rightarrow$}
\left\{\!\begin{aligned}
\ket{3}\ket{2}\ket{\xi_\pm},~\ket{\eta_\pm}\ket{2}\ket{3},~\tri{0}{3}{\chi_\pm},~\tri{2}{\chi_\pm}{2},~\tri{\chi_\pm}{3}{1},~~~\\
|3\rangle|2\rangle|3\rangle,~|3\rangle|3\rangle|0\rangle,~|3\rangle|3\rangle|2\rangle,~|3\rangle|3\rangle|3\rangle,~
|2\rangle|2\rangle|0\rangle,~|2\rangle|2\rangle|1\rangle,~|2\rangle|3\rangle|0\rangle,\\
|2\rangle|3\rangle|3\rangle,~|0\rangle|2\rangle|0\rangle,~|0\rangle|2\rangle|1\rangle,~|0\rangle|2\rangle|2\rangle,~
|0\rangle|3\rangle|0\rangle,~|0\rangle|3\rangle|1\rangle,~|1\rangle|2\rangle|0\rangle,\\
|1\rangle|2\rangle|1\rangle,~|1\rangle|2\rangle|2\rangle,~|1\rangle|3\rangle|0\rangle,~|1\rangle|3\rangle|1\rangle,~|1\rangle|3\rangle|2\rangle,~|1\rangle|3\rangle|3\rangle~~~~~~~~~~~~
\end{aligned}\right\}~~~~~,	\\\nonumber\\
\rotatebox[origin=c]{0}{$\mbox{Entangled}\Rightarrow$}
\left\{\!\begin{aligned}
\ket{3}_A\left(\ket{1}_B\ket{000}_{abc}\pm\ket{2}_B\ket{111}_{abc}\right)\ket{0}_C,\\ 
\ket{2}_A\left(\ket{1}_B\ket{000}_{abc}\pm\ket{2}_B\ket{111}_{abc}\right)\ket{3}_C
\end{aligned}\right\}.~~~~~~~~~~~~~~~~~~~~~~~~~~~~~~~~~~~~~~~~~~~~~
\end{align}

{\bf Step-2:} Alice performs the measurement 
$$\mathcal{K}\equiv\left\lbrace K_1:=\mathbb{P}\left[\ket{0}_A;\ket{0}_{a}\right],~K_2:=\mathbb{P}\left[\left(\ket{0},\ket{1}\right)_A;\ket{1}_{a}\right] ,~K_3:=\mathbb{I}-K_1-K_2\right\rbrace .$$
States corresponding to the outcomes $K_1$ and $K_2$ are listed below:
\begin{align}\label{c13}
\rotatebox[origin=c]{0}{$K_1\Rightarrow$}
\left\{\!\begin{aligned}
\bi{0}{\eta_\pm}\ket{3},~|0\rangle|0\rangle|0\rangle,~|0\rangle|0\rangle|1\rangle,~|0\rangle|0\rangle|2\rangle,~|0\rangle|1\rangle|0\rangle,
~|0\rangle|1\rangle|1\rangle,~|0\rangle|1\rangle|2\rangle
\end{aligned}\right\},~~~~~~~\\\nonumber\\
\rotatebox[origin=c]{0}{$K_2\Rightarrow$}
\left\{\!\begin{aligned}
\ket{\eta_\pm}\ket{2}\ket{3},~\tri{0}{3}{\chi_\pm},~|0\rangle|2\rangle|0\rangle,~|0\rangle|2\rangle|1\rangle,~|0\rangle|2\rangle|2\rangle,~
|0\rangle|3\rangle|0\rangle,~|0\rangle|3\rangle|1\rangle,\\
|1\rangle|2\rangle|0\rangle,~|1\rangle|2\rangle|1\rangle,~|1\rangle|2\rangle|2\rangle,~|1\rangle|3\rangle|0\rangle,~|1\rangle|3\rangle|1\rangle,~|1\rangle|3\rangle|2\rangle,~|1\rangle|3\rangle|3\rangle~~~~~~
\end{aligned}\right\}.
\end{align}
Both of these two sets are LOCC distinguishable. If $K_3$ clicks the state is one of the remaining $40$ states. 

{\bf Step-3:} Charlie performs the measurement 
$$\mathcal{N}\equiv\left\lbrace N_1:=\mathbb{P}\left[\ket{2}_C;\ket{0}_{c}\right],~N_2:=\mathbb{P}\left[\left(\ket{1},\ket{2}\right)_C;\ket{1}_{c}\right] ,~N_3:=\mathbb{I}-N_1-N_2\right\rbrace .$$
States corresponding to the outcomes $N_1$ and $N_2$ are listed below:
\begin{align}\label{c14}
\left\{\!\begin{aligned}
N_1\Rightarrow&\left\lbrace \ket{3}\bi{\eta_\pm}{2},~|2\rangle|0\rangle|2\rangle,~|2\rangle|1\rangle|2\rangle,~|1\rangle|0\rangle|2\rangle,~|1\rangle|1\rangle|2\rangle\right\rbrace,\\
N_2\Rightarrow&\left\lbrace \ket{3}\ket{2}\ket{\xi_\pm},~\tri{2}{\chi_\pm}{2},~\tri{\chi_\pm}{3}{1},~|3\rangle|3\rangle|2\rangle,~|2\rangle|2\rangle|1\rangle\right\rbrace 
\end{aligned}\right\}.
\end{align}
The states corresponding to outcome $N_1$ are LOCC distinguishable. LOCC distinguishability of the set of states corresponding to $N_2$ is discussed later (see Remark \ref{remark1}). If $N_3$ clicks the given state is one of the remaining $26$ states.

{\bf Step-4:} Bob performs the measurement 
$$\mathcal{M}^\prime\equiv\left\lbrace M^\prime_1:=\mathbb{P}\left[\ket{0}_B;\mathbb{I}_{b}\right],~M^\prime_2:=\mathbb{P}\left[\ket{3}_B;\mathbb{I}_{b}\right] ,~M^\prime_3:=\mathbb{I}-M^\prime_1-M^\prime_2\right\rbrace .$$
States corresponding to the outcomes $M^\prime_1$ and $M^\prime_2$ are listed below:
\begin{align}\label{c15}
\left\{\!\begin{aligned}
M^\prime_1\Rightarrow&\left\lbrace \tri{3}{0}{\eta_\pm},~\tri{\xi_\pm}{0}{3},~\tri{3}{0}{3},~\tri{2}{0}{0},~\tri{2}{0}{1},~\tri{1}{0}{0},~\tri{1}{0}{1}\right\rbrace,\\
M^\prime_2\Rightarrow&\left\lbrace \tri{3}{3}{0},~\tri{3}{3}{3},~\tri{2}{3}{0},~\tri{2}{3}{3}\right\rbrace 
\end{aligned}\right\}.
\end{align}
Evidently, these two sets are LOCC distinguishable. If $M^\prime_3$ clicks, given is one of the remaining $13$ states.

{\bf Step-5:} Alice performs the measurement 
$$\mathcal{K}^\prime\equiv\left\lbrace K^\prime_1:=\mathbb{P}\left[\ket{1}_A;\mathbb{I}_{a}\right],~K^\prime_2:=\mathbb{P}\left[\ket{2}_A;\mathbb{I}_{a}\right] ,~K^\prime_3:=\mathbb{I}-K^\prime_1-K^\prime_2\right\rbrace .$$
States corresponding to the outcomes are listed below:
\begin{align}\label{c16}
\left\{\!\begin{aligned}
K^\prime_1\Rightarrow&\left\lbrace \tri{1}{1}{3},~\tri{1}{1}{0},~\tri{1}{1}{1}\right\rbrace,\\
K^\prime_2\Rightarrow&\left\lbrace \tri{2}{1}{0},~\tri{2}{1}{1},~\tri{2}{2}{0},~\ket{2}_A\left(\ket{1}_B\ket{000}_{abc}\pm\ket{2}_B\ket{111}_{abc}\pm\right)\ket{3}_C\right\rbrace ,\\
K^\prime_3\Rightarrow&\left\lbrace \tri{3}{1}{1},~\tri{3}{1}{3},~\tri{3}{2}{3},~\ket{3}_A\left(\ket{1}_B\ket{000}_{abc}\pm\ket{2}_B\ket{111}_{abc}\pm\right)\ket{0}_C\right\rbrace 
\end{aligned}\right\}.
\end{align}
Evidently all these three sets are LOCC distinguishable.

\begin{remark}\label{remark1}
If $N_2$ clicks in {\bf Step-3} then the given state is one of 
\begin{equation}
\left\lbrace \ket{3}\ket{2}\ket{\xi_\pm},~\tri{2}{\chi_\pm}{2},~\tri{\chi_\pm}{3}{1},~|3\rangle|3\rangle|2\rangle,~|2\rangle|2\rangle|1\rangle\right\rbrace.
\end{equation}
Considering the relabeling $2\mapsto1$, $3\mapsto0$ for Alice \& Bob and $1\mapsto1$, $2\mapsto0$ for Charlie, the above set reads as,  
\begin{equation}
\left\lbrace \tri{0}{1}{0\pm 1},~\tri{1}{0\pm 1}{0},~\tri{0\pm 1}{0}{1},~\tri{0}{0}{0},~\tri{1}{1}{1}\right\rbrace.
\end{equation}
It is the OPB corresponding to the Shift UPB of $(\mathbb{C}^2)^{\otimes3}$ \cite{Bennett99}, and this set can be perfectly distinguished under LOCC if a $2$-qubit maximally entangled state is shared between any two parties. Since the unknown state is chosen at random ({\it i.e} with uniform probability) from the set $\mathbb{B}_{II}(4,3)$, therefore total entanglement consumed in this protocol are $1$ $\mbox{GHZ}$ and $\frac{1}{8}$ $\mbox{EPR}$ .   
\end{remark}

\begin{remark}\label{remark2}
Instead of GHZ resource, consider that Alice and Bob share a $2$-qubit maximally entangled state. After {\bf Step-1}, the tag is shared between Alice and Bob only. As already discussed in {\bf Step-2} Alice can discriminate $20$ state corresponding to the outcomes $K_1$ and $K_2$. However, if $K_3$ outcome occurs, the discrimination protocol can not be further proceeded if no more entangled resource is used. But if an entangled state $\ket{\phi^+}_{b^\prime c}$ is provided between Bob and Charlie then a perfect discrimination protocol is possible. For that Bob starts with a twist-breaking measurement $\{\mathbb{P}[(\ket{1},\ket{2})_B;\ket{0}_{b^\prime}]+\mathbb{P}[(\ket{0},\ket{3})_B;\ket{1}_{b^\prime}],~\mathbb{I}-\mathbb{P}\}$. Then an analogous protocol follows that discriminate all the remaining $44$ states. Therefore total entanglement consumption in this protocol is $\left\lbrace(1,\ket{\phi^+}_{\mathcal{AB}});(\frac{11}{16},\ket{\phi^+}_{\mathcal{BC}}) \right\rbrace $.       
\end{remark}

\twocolumngrid

\end{document}